\documentclass[twocolumn,secnumarabic,amssymb,showpacs, nobibnotes, aps, prl,floatfix]{revtex4-1}
\usepackage{graphicx}
\begin{document}

\title{Stabilization of periodically forced Hele-Shaw flows by means of a non-monotonous viscosity profile}

\author{Vicente P\'erez-Mu\~nuzuri}
\email[Email:~]{vicente.perez.munuzuri@usc.es}
\affiliation{CRETUS. Group of Nonlinear Physics, Faculty of Physics, University of Santiago de Compostela. E-15782 Santiago de Compostela, Spain}
\date{\today}

\begin{abstract}
The onset of viscous fingering in the presence of a viscosity profile is investigated theoretically for two immiscible fluids undergoing a time-dependent injection. Here, we show that the presence of a positive viscosity gradient at the interface between both fluids stabilizes the interface facilitating the spread of the perturbation. This effect is much more pronounced in the case of sinusoidal injection flows. The influence of the viscosity gradient on the dispersion relation is analyzed.
Numerical simulations of the Navier-Stokes equation confirm the linear stability analysis.
\end{abstract}



\maketitle
\section{Introduction}

Saffman-Taylor instability \cite{Taylor58} may arise when two fluids of different viscosities are pushed by a pressure gradient through two horizontal parallel plates (Hele-Shaw (HS) cell). It is well known, both experimentally and theoretically, that when a less viscous Newtonian fluid displaces a more viscous one, a fingering instability develops at the interface between both immiscible fluids \cite{Faber95}. For non Newtonian fluids, an unexpected propagation of fractures may develop in the invaded fluid for specific conditions \cite{Nittmann85,Wilson90,Mora}.

Due to chemical and oil recovery applications, the possibility to control this interfacial instability has been the subject of numerous studies in the last decades \cite{Araktingi93}. Chemical reactions occurring at the interface drive viscosity changes that modulate the hydrodynamic viscous fingering instability, both for unstable \cite{Nagatsu,DeWit20,Escala21,Tafur21} or stable \cite{Trevelyan11} fronts. For the last case, for example, when a change in pH is chemically induced at the interface, some fingers may develop \cite{Riolfo12,Escala19}. Without using a chemical reaction, the process involves displacing the most viscous fluid first with some polymer-thickened fluid (usually an aqueous phase containing water and polymer in appropriate proportion to have the desired profile) followed by the less viscous one. This strategy basically involves a three-layer fluid with an intermediate layer of finite thickness containing the polymer solution \cite{Daripa}. Other possibilities for viscous fingering in the presence of non-monotonic viscous profiles have also been studied \cite{Manickam93,Haudin16}. Other non-chemical strategies to control this instability include, imposing a time-dependent injection rate, tapering the gap between plates so they are no longer parallel \cite{Dias13,Zheng15}, replacing one of the plates with an elastic membrane \cite{Pihler18}, rotating the HS cell while the inviscid fluid is injected \cite{Morrow19}, or competition between gravity and viscous forces \cite{Ruith00}, among others.

Experiments and numerical simulations involving the injection of a less viscous fluid into a more viscous one in a radial HS cell showed that stronger injection rates result in an enhancement of viscous fingering. Most of the existing studies have focused on displacements under constant injection. There are however numerous practical processes where the injection is actually time dependent \cite{Barrio07,Torralba08,Gosayir13,Lins16,Morrow19,Coutinho20}. Most of these studies showed that better recovery can be achieved through time-dependent injection schemes in comparison with constant injection ones due to an attenuation of the finger instability. On the other hand, several authors \cite{Lins17} demonstrated that for a periodic injection on a radial HS cell the number of fingers and their structures could be controlled. As opposite, other authors have shown that fingering destabilization due to periodic injection favors fluid mixing in confined systems, such as in microfluidic devices \cite{mixing}.

Based on the results reported, it is clear that injection rates and periodicity play an important role to control the flow behavior and interface instability. The objective of the present study is to analyze the combination of imposing a time-dependent injection rate with the presence of non monotonic viscosity profiles at the interface. While we observed that the periodic injection leads to a resonance-like effect destabilizing the interface, the presence of a three-layer non-monotonic viscosity profile stabilize the front achieving better recovery results for the viscous fluid. In this work, we present a linear stability analysis for the onset of viscous fingering for two immiscible fluids subject to a periodic injection and a viscosity gradient at the interface. Viscous fingering has been observed numerically confirming this analysis and the dependence with parameters of the model studied.


\section{Linear stability analysis}

The equation of motion of an incompressible Newtonian fluid is given by,
\begin{eqnarray}
\rho\frac{Du_i}{Dt}&=&-\frac{\partial p}{\partial x_i} + \frac{\partial}{\partial x_j}\left(2\mu e_{ij}\right), \label{eq:NS1} \\
\frac{\partial u_i}{\partial x_i} &=& 0
\label{eq:div}
\end{eqnarray}
where $D/Dt$ is the material derivative, $\rho$ is the constant density, $u_i$ the velocity field, $p$ the pressure, $\mu$ the dynamic viscosity, and $e_{ij}=(\partial u_i/\partial x_j+\partial u_j/\partial x_i)/2$ is the strain rate tensor.
For a non monotonic viscosity profile $\mu=\mu(x_i)$, Eq.~(\ref{eq:NS1}) becomes,
\begin{equation}
\rho\frac{Du_i}{Dt}=-\frac{\partial p}{\partial x_i} + \mu\frac{\partial^2u_i}{\partial x_j \partial x_j} + 2e_{ij}\frac{\partial \mu}{\partial x_j}
\label{eq:NS2}
\end{equation}

\begin{figure}
\includegraphics[width=1.\linewidth]{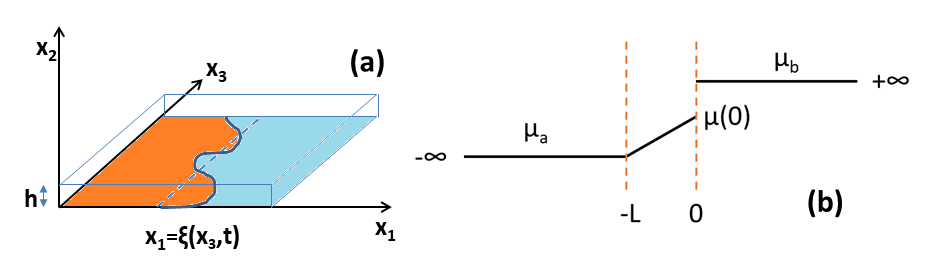}
\caption[]{Sketch of the Hele-Shaw cell (a) and the piecewise viscosity profile $\mu(x_1)$ (b) used here. $L$ stands for the region where a linear viscosity gradient is considered between both fluids with constant viscosities $\mu_a$ and $\mu_b$.}
\label{fig:setup}
\end{figure}

We consider the case where a viscous fluid with viscosity $\mu_a$ is pushing a more viscous one $\mu_b$ in the $x_1$-direction between closely spaced parallel plates separated some distance $h$ as it is shown in Fig.~\ref{fig:setup}(a), subject to a perturbation at the interface $x_1=\xi(x_3,t)$. In between both fluids a region with a viscosity gradient is considered, Fig.~\ref{fig:setup}(b). 
In the basic HS flow, we suppose a constant negative pressure gradient along the $x_1$ axis so that the flow goes from the left ($x_1<0$) to the right ($x_1>0$) with velocity field $[u_1^0(x_2),0,0]$.
Following the standard decomposition in normal modes, we perturb the system of equations (\ref{eq:NS2}). Thus, the perturbed velocity field is $[u_1^0(x_2)+\epsilon u_1^1(x_i,t),0,\epsilon u_3^1(x_i,t)]$, the pressure $P^0+\epsilon P^1(x_i,t)$, and the interface equation is $x_1=\xi^0(t)+\epsilon \xi^1(x_3,t)$ ($\epsilon \ll 1$). Assuming a sinusoidal perturbation along the $x_3$ axis, the perturbed quantities can be written as,
\begin{eqnarray}
\xi^1(x_3,t)&=& \xi\exp(\imath k_3x_3 + \omega t) \nonumber\\
u_1^1(x_i,t)&=& u_{a,b}(x_2)\exp(\imath k_3x_3 + \omega t)\exp\left(\pm k_1x_1\right) \nonumber\\
u_3^1(x_i,t)&=& w_{a,b}(x_2)\exp(\imath k_3x_3 + \omega t)\exp\left(\pm k_1x_1\right) \nonumber\\
P^1(x_1,x_3,t)&=& P_{a,b}\exp(\imath k_3x_3 + \omega t)\exp\left(\pm k_1x_1\right) \nonumber
\end{eqnarray}
where $\pm$ stands for the left (a) and right (b) fluids with viscosities $\mu_a$ and $\mu_b$, respectively. $\xi$, $u(x_2)$, $w(x_2)$, and $P$ are the amplitudes of the normal modes of the perturbation. The problem is completed by the no-slip boundary condition at the plates $u_i=0$ for $x_2=0,h$. Without loss of generality, from now on, we assume a piecewise viscosity profile $\mu=\mu(x_1)$ as depicted in Fig.~\ref{fig:setup}(b).

Rewriting Eq.~(\ref{eq:NS2}) at zero order we obtain,
\begin{equation}
u_1^0(x_2) = \frac{G}{2\mu_{a,b}}\left(x_2^2-hx_2\right)
\label{eq:Darcy}
\end{equation}
with $G$ the negative pressure gradient along the $x_1$ axis. Averaging in the $x_2$ direction, the Darcy's law $U=\langle u_1^0\rangle = -Gh^2/12\mu_{a,b}$ is obtained.
The interface between the two fluids is $x_1=\xi^0(t)$ with $\xi^0(t)=Ut$.

At first order, the amplitudes of the normal modes $u_{a,b}$ are given by,
\begin{equation}
\rho\left(\omega u_{a,b} \pm Uku_{a,b}\right) = \mp P_{a,b} - \frac{12\mu_{a,b}}{h^2}u_{a,b} \pm 2\mu'_{a,b}(x)ku_{a,b}
\label{eq:u}
\end{equation}
with $\mu'_{a,b}(x)=\partial\mu_{a,b}/\partial x$, and we have used the incompressibility of the flow to show that $k=k_1=k_3$. A similar equation can be obtained for the mode $w_{a,b}$.

Three conditions must be imposed at the interface, namely the kinematical condition and the condition of continuity of normal and tangential stresses,
\begin{eqnarray}
u_1^1(x_2)_{a,b} &=& \frac{\partial \xi^1}{\partial t}, \nonumber\\
\left(P^0_a+P^1_a\right) - \left(P^0_b + P^1_b\right) &=& -\gamma \frac{\partial^2\xi^1}{\partial x_3^2}, \label{eq:bound} \\
\mu_a \frac{\partial u_a}{\partial x_2} &=& \mu_b \frac{\partial u_b}{\partial x_2} \nonumber
\end{eqnarray}
where $\gamma$ is the surface tension at the interface, and $P^0$ is obtained integrating the Darcy's law.

\begin{figure}
\includegraphics[width=1.\linewidth]{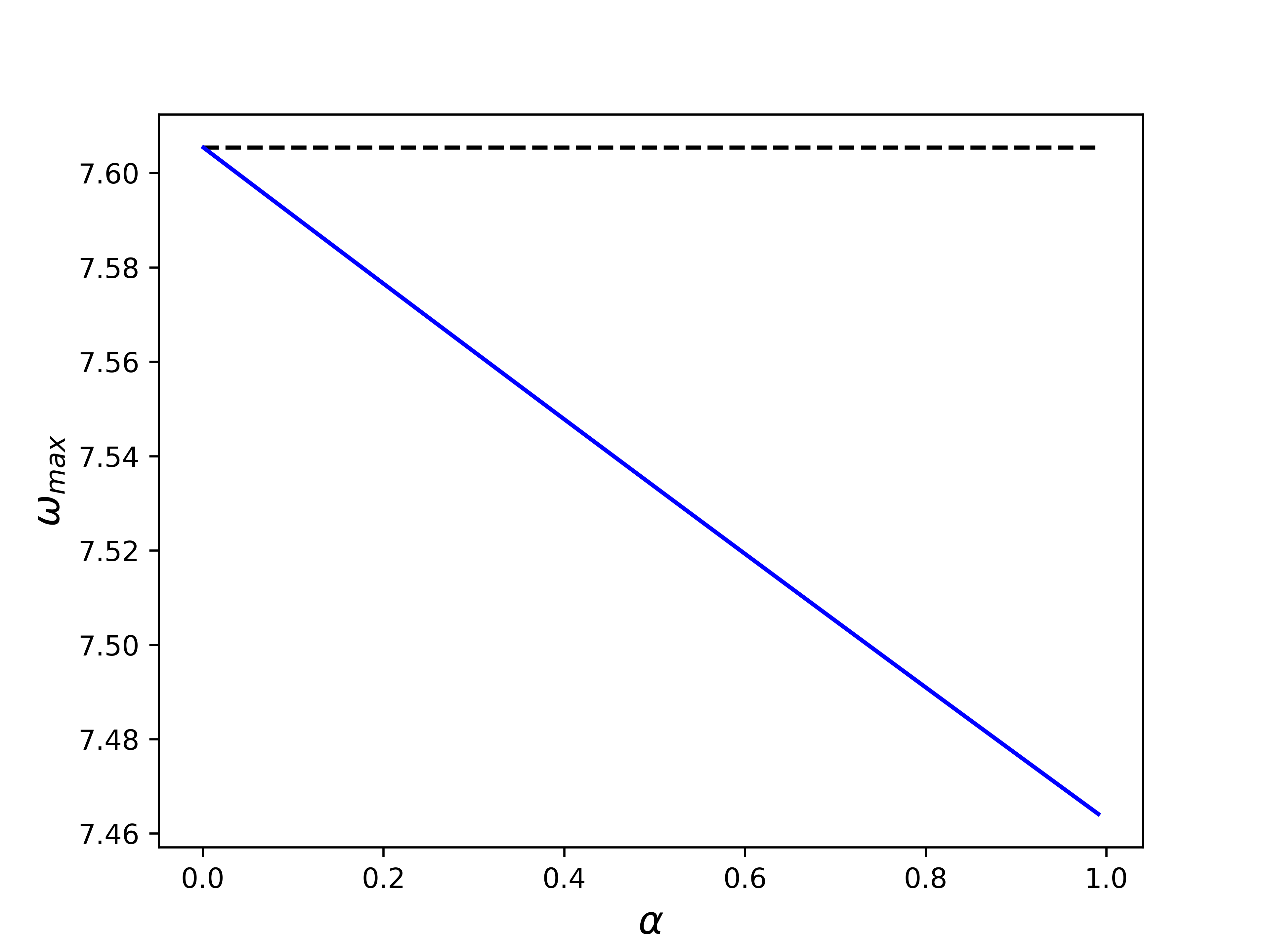}
\caption[]{Maximum growth rate $\omega_{max}$ as a function of $\alpha$, $\mu'_a=(\alpha\mu_b-\mu_a)/L$, for the unstable case ($\mu_b>\mu_a$). Dashed line corresponds to the Saffman-Taylor case. Set of parameters: $h=10^{-3}$ m, $U=0.01$ m/s, water $\mu_a=0.896\cdot 10^{-3}$ Pa\,s, glycerine $\mu_b=0.95$ Pa\,s, $\gamma=29.06\cdot 10^{-3}$ N/m, $L=0.01$, and $\mu'_b=0$}
\label{fig:dispersion}
\end{figure}

Solving analytically the above equations, we obtain the dispersion relation,
\begin{equation}
\omega(k) = \frac{Uk(\mu_b-\mu_a) - \frac{\gamma h^2}{12} k^3}{\mu_a+\mu_b + \frac{k h^2}{6}(\mu'_a-\mu'_b)}
\label{eq:omega}
\end{equation}
Note that the classical Saffman-Taylor dispersion relation ($\mu_b>\mu_a$) \cite{Faber95} is recovered for $\mu'_{a,b}=0$. An improvement of the stability (lower values of $\omega(k)$) is obtained for $\mu'_a-\mu'_b>0$ compared to the Saffman-Taylor case. Fig.~\ref{fig:dispersion} shows this improvement as the viscosity gradient increases for $\mu'_a=(\alpha\mu_b-\mu_a)/L$ ($\alpha=\mu(0)/\mu_b$) and $\mu'_b=0$.

\section{Numerical simulations}
Two-dimensional ($x_1,x_3$) direct numerical simulations (DNS) of the Navier-Stokes equations (\ref{eq:NS2}) have been used with boundary conditions $u_i=0$ for $x_3=0,x_3^{max}$, and $\partial_1 u_i=0$ for $x_1=0,x_1^{max}$. At the interface between both fluids, the kinematical condition and the condition of continuity of normal and tangential stresses hold. Eqs.~(\ref{eq:NS2}) were integrated using an implicit Crank-Nicholson method for the advection and diffusion terms.

Pressure is obtained by means of a fractional-step, or time-splitting scheme \cite{Chorin68,Moin85}. 
Process of solving the momentum and the pressure Poisson equations is repeated until the velocity field is divergence free at each instant of time. The pressure gradient is assumed to oscillate as $\partial_1 P=G+P'\sin(\Omega t)$ with $G$ the negative pressure gradient along the $x_1$-axis. The pressure boundary conditions are of Neumann type on $x_3=0,x_3^{max}$.


The assumed rectangular computational domain is discretized on a staggered Cartesian grid when pressure $P$ and velocities are determined on three grids shifted relative to each other. So, $P$ is located in the center of each cell, the $x_1$-component velocity $u_1$ is on the middle points of vertical faces, and the $x_3$-component velocity $u_3$ is on the middle points of horizontal faces. 

The convergence of the code was first tested by varying the spatial resolution from $1000\times 100$ to $4000\times 400$. Accordingly, the time step was changed from 0.001 to 0.1, and nearly the same time
evolution of fingers were obtained. To ensure a good numerical convergence and relatively shorter calculation time, a spatial resolution of $2000\times 200$ and a time step $\Delta t = 0.01$ were adopted for the range of parameters examined throughout this study.

Initially we suppose an inlet front of viscosity $\mu_a$ pushing periodically another fluid with viscosity $\mu_b >\mu_a$ and the interface between both fluids has an initial sinusoidal perturbation (amplitude 20 and wavenumber $\pi/50$) and a viscosity gradient $\mu'_a$.

\section{Results}

\begin{figure}
\includegraphics[width=1.\linewidth]{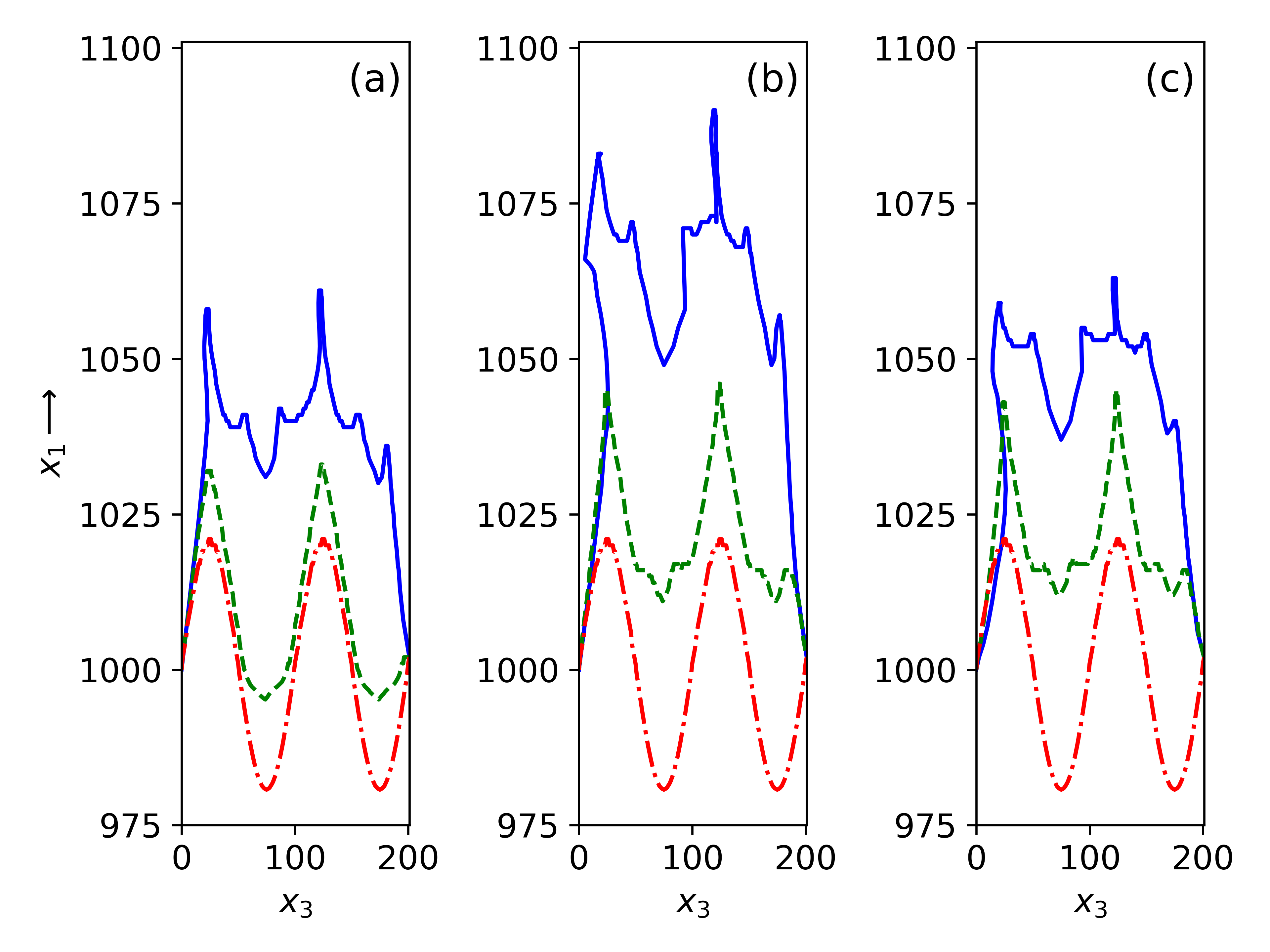}
\caption[]{Interface fronts between fluids for three forcing frequencies $\Omega$, with (blue and green dashed lines) and without periodic forcing (red dot-dashed line). Green dashed interface corresponds to an intermediate time. $\alpha=0.4$, $G=-1$ Pa/m, $P'=1$ Pa/m and rest of parameters as in Fig.~\ref{fig:dispersion}. From left to right, $\Omega=0.004,0.012,0.026$, respectively.}
\label{fig:frentes}
\end{figure}

\begin{figure}
\includegraphics[width=1.\linewidth]{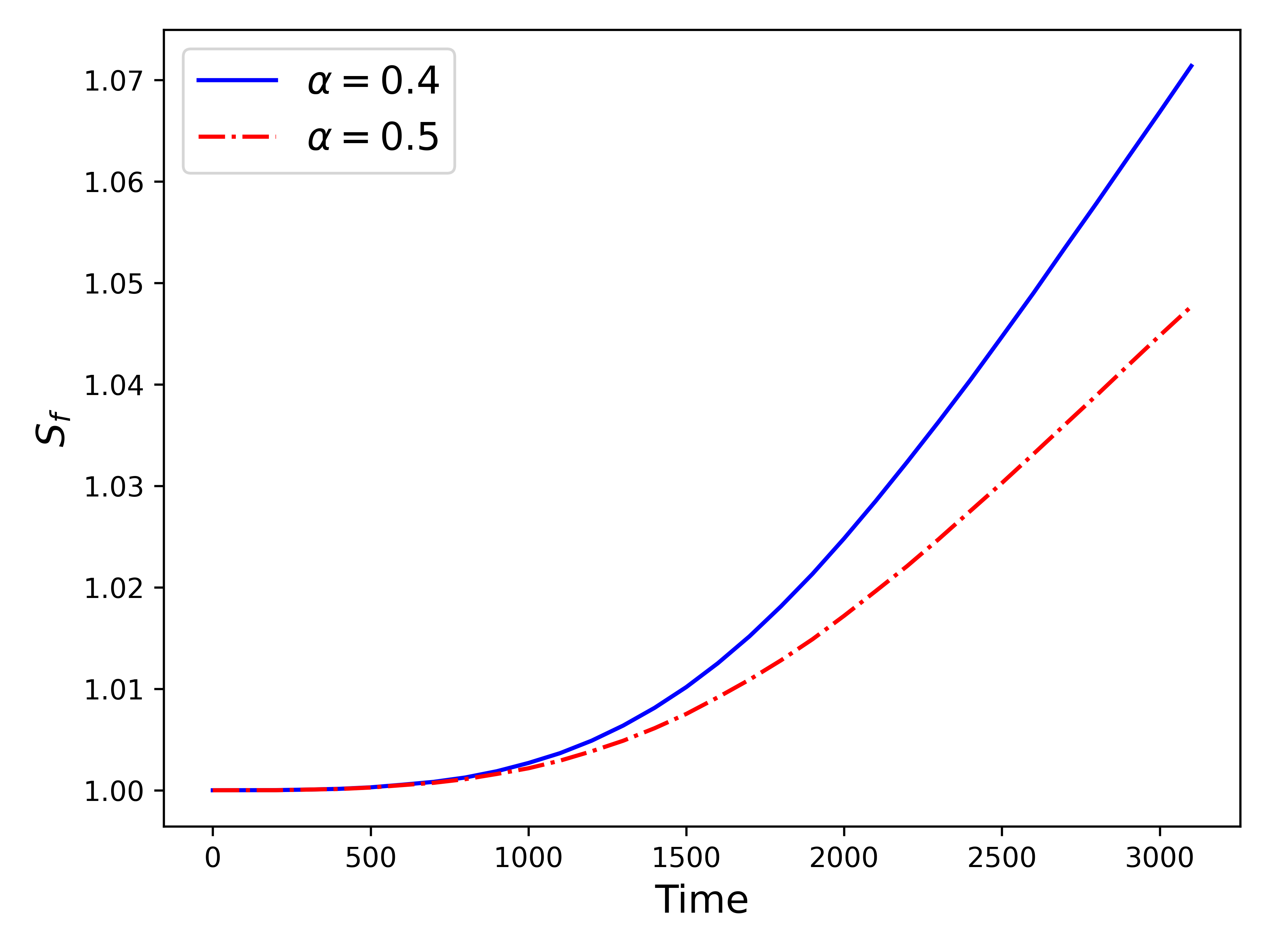}
\caption[]{Time evolution of the surface ratio $S_f$ for two values of the ratio $\alpha=\mu(0)/\mu_b$. $\Omega=0.012$. }
\label{fig:evolucion}
\end{figure}

It is well known that for HS flows, under unstable conditions ($\mu_b>\mu_a$) the interface between both fluids becomes unstable and some fingers develop. This process is enhanced under a pulsating inlet flow. Figure~\ref{fig:frentes} shows, for two instants of time, the fluids interface for three different forcing frequencies $\Omega$ compared to the case without forcing. Note that in all cases the interface has advanced faster under forcing, and the net displacement depends on the forcing frequency. For an intermediate value of $\Omega$ the front advances faster than for the other two frequencies.
The ratio of the surfaces $S_f$ covered by the flow interface, with and without forcing, allows us to quantify the effect of forcing, and can be considered proportional to the ratio between the linear growth rates for both cases. Figure~\ref{fig:evolucion} shows the time evolution of $S_f$ for two values of $\alpha$ for the intermediate frequency shown in Fig.~\ref{fig:frentes}. Note that the front accelerates with time, faster as $\alpha$ diminishes. 

\begin{figure}
\includegraphics[width=1.\linewidth]{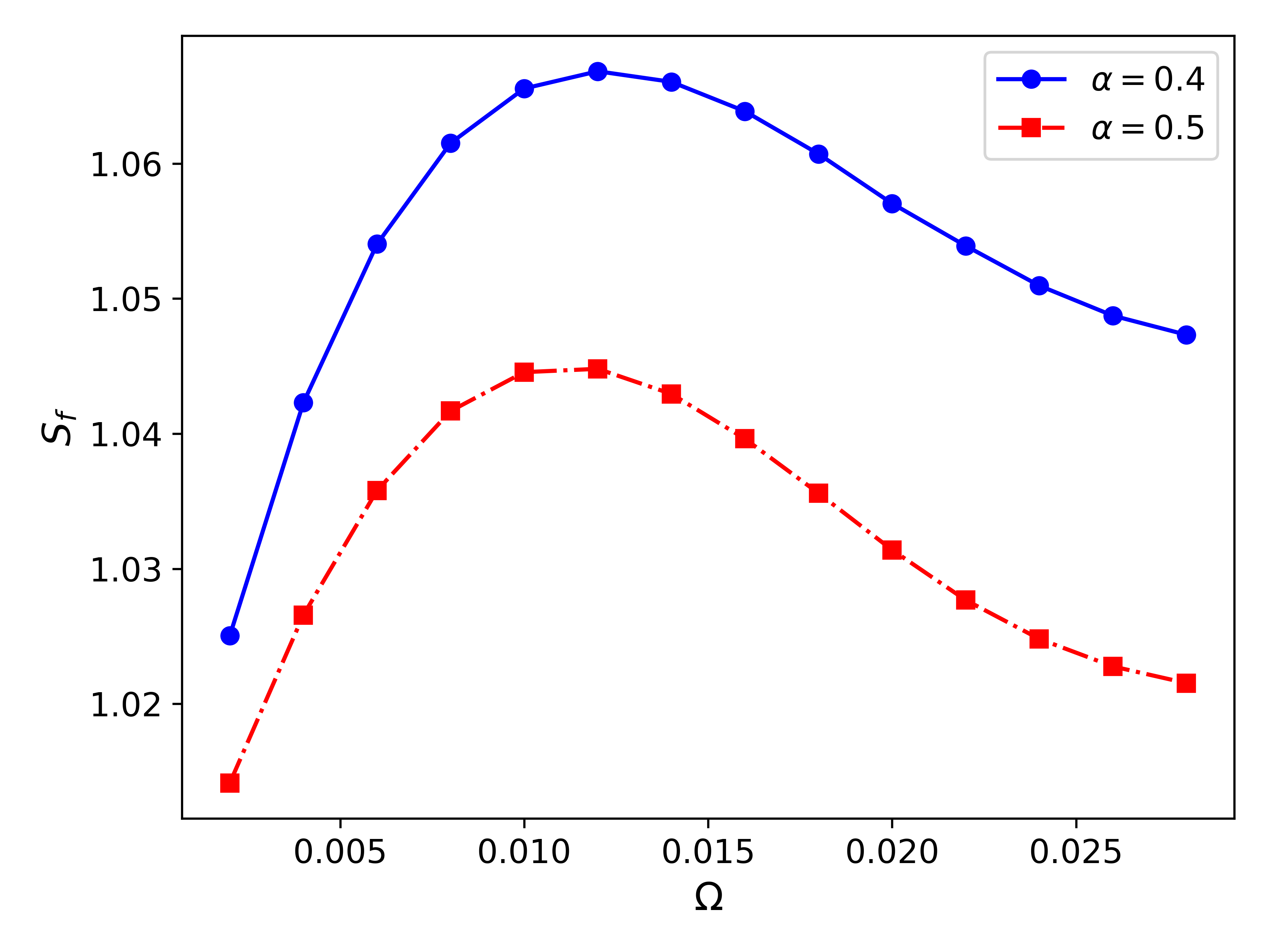}
\caption[]{Surface ratio $S_f$ covered by the fluid interfaces, with and without periodic forcing, as a function of the forcing frequency $\Omega$ for two values of the ratio $\alpha=\mu(0)/\mu_b$. Parameters as in Fig.~\ref{fig:dispersion}.}
\label{fig:resonancia}
\end{figure}

\begin{figure}
\includegraphics[width=1.\linewidth]{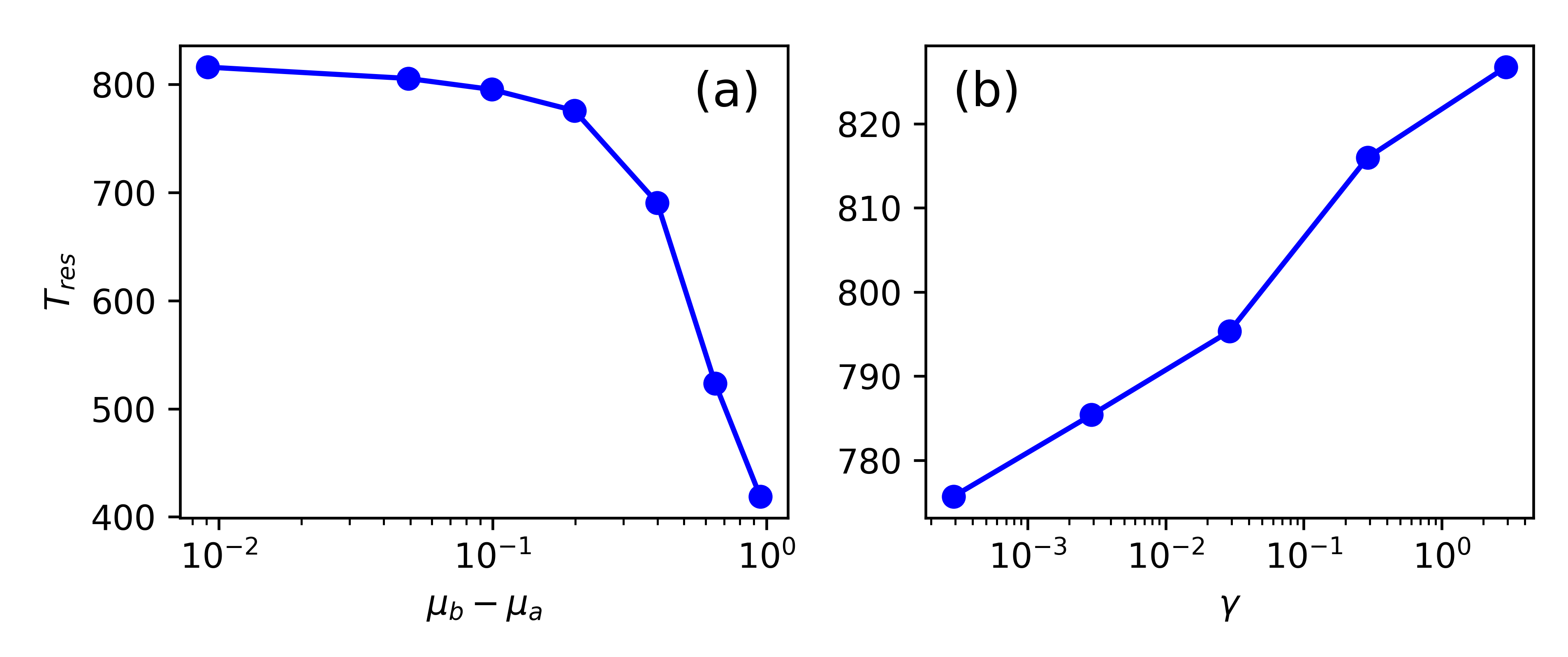}
\caption[]{Resonant period $T_{res}$ as a function of the viscosity (a) and the surface tension $\gamma$ (b). In all cases, the behavior of $T_{res}$ coincides with the characteristic time of the flow $t_{ch}\propto \gamma/(\mu_b-\mu_a)$. $\gamma=29.06\cdot 10^{-3}$ N/m (a) and $\mu_b=0.2$ Pa\,s (b). $\alpha=0.4$ and $\mu_a=0.89\cdot 10^{-3}$ Pa\,s in all cases.}
\label{fig:area_max}
\end{figure}

Figure~\ref{fig:resonancia} shows the ratio $S_f$ as a function of the forcing frequency $\Omega$ for the same instant of time as in Fig.~\ref{fig:frentes}. Note the significant enhancement of the flow instability at certain intermediate frequencies that was clearly visible in Fig.~\ref{fig:frentes}. For very small and large forcing frequencies, $S_f\rightarrow S_f(P'=0)=1$. Maximum $S_f$ value was attained always for the same resonant frequency $\Omega_{res}$ for any value of the ratio $\alpha=\mu(0)/\mu_b$, and seems to be the signature of a resonant behavior between the characteristic time scale of the flow, ratio of the surface tension and the viscous (pressure drop) forces,
\begin{equation}
t_{ch}\equiv\frac{\gamma}{U\Delta P} \propto \frac{\gamma}{\mu_b-\mu_a}
\label{eq:t_ch}
\end{equation}
and the forcing frequency. A further investigation of this phenomenon was carried out for other values of viscosity $\mu_b$ and surface tension $\gamma$. Changing these parameters modifies the structure of the interface (for example, increasing $\gamma$ leads to a smoothing of the front), however, the phenomenon of resonance was found in all examined cases, although a change in the resonant period $T_{res}$ was observed (Fig.~\ref{fig:area_max}) consistent with $t_{ch}$ (\ref{eq:t_ch}).
The previous results show that at the resonant frequency, the interface can be characterized by an increased instability, which appears to emerge earlier and be stronger than for the other flow periods.

\begin{figure}
\includegraphics[width=1.\linewidth]{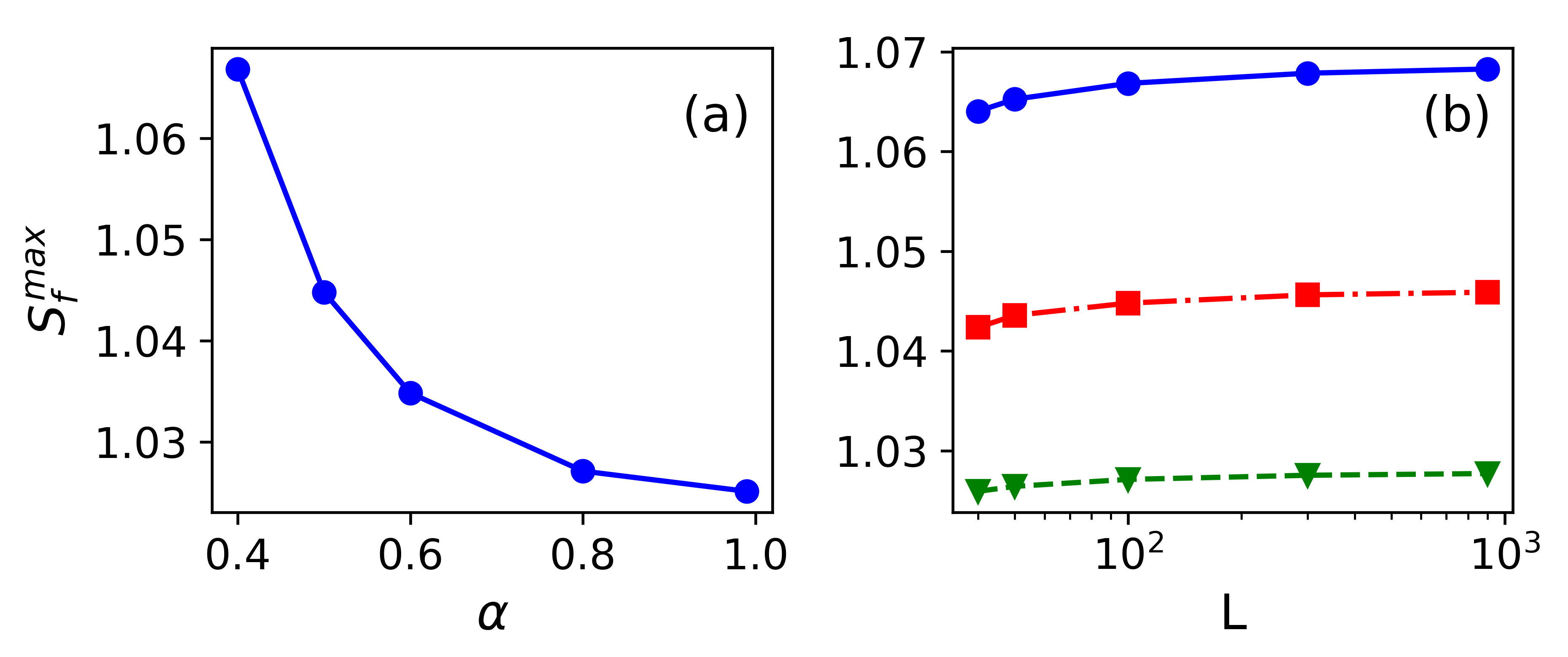}
\caption[]{Maximum surface ratio $S_f$ as a function of the ratio $\alpha=\mu(0)/\mu_b$ (a) and the length $L$ of the middle layer (b). In panel (b) lines correspond to $\alpha=0.4$ (circles and blue solid line), $\alpha=0.5$ (squares and dot-dashed red line) and $\alpha=0.8$ (triangles and dashed green line). Parameters as in Fig.~\ref{fig:dispersion}.}
\label{fig:grad}
\end{figure}

The dependence of the flow instability enhancement at the resonance peak on the values of $\alpha$ and the length of the middle layer $L$ is shown in Fig.~\ref{fig:grad}. Note that increasing $p$ ($L$), increases (decreases) the viscosity gradient $\mu'_a=(\alpha\mu_b-\mu_a)/L$ at the interface, and $S_f$ diminishes (increases) as expected from the theoretical dispersion relation, Eq.~(\ref{eq:omega}) (Fig.~\ref{fig:dispersion}). Increasing the gradient leads to stabilize the front diminishing the growth rate of instabilities at the interface. 
For $\alpha\rightarrow \mu_a/\mu_b$ the classical two-layer Saffman-Taylor instability is obtained but the risk of interface breakup increases. 
For smaller ratios $\mu_b/\mu_a$ than those used here, it is possible to reach that limit without the interface eventually breaks up. Note in Fig.~\ref{fig:frentes} the formation of plumes or spikes when periodic forcing is applied. When no periodic forcing is considered, the initial sinusoidal perturbation evolves into a single finger and no cusps were observed. These plumes smooth out as surface tension increases. As $\alpha$ diminishes, the interface moves faster, increasing inertial forces compared to surface tension (large Weber number), favouring the formation of elongated plumes and the interface may break up as it has been observed in other interfacial instabilities \cite{interface}. Thus, we may assume that periodic forcing increases inertial forces as the interface moves faster than without forcing, and cusps/plumes appear before the interface breaks up. Once this happens our numerical algorithm is not stable. Different numerical approaches should then be undertaken but they are beyond the scope of this paper.

\section{Conclusions}

The onset condition of viscous fingering for a fluid displacing a more viscous one in a Hele-Shaw cell undergoing time-dependent injection in a homogeneous medium has been studied in the presence of a non monotonic viscosity profile in the direction of motion. For wave numbers above a critical one, perturbations at both sides of the interface spread in the same direction, destabilizing the interface. The spreading velocity is modulated by the term $\exp{(\omega t)}$ that broads the fingering in the direction of motion. The growth instability rate is smaller when the viscosity profile has a positive slope on the interface, Eq.~(\ref{eq:omega}). Periodic forcing enhances formation of a single finger displacing the more viscous fluid, while the presence of an intermediate layer with a positive viscous gradient stabilizes the interface preventing breakage. 

For a critical value of the injection period the interface between both fluids grows significantly if compared to other examined periods. This critical value of the period is suspected to be associated with a resonance-like dynamics between the external injection forcing and the characteristic time scale of the flow (ratio of the surface tension and the viscous (pressure drop) forces). The effect of a viscous linear gradient at the interface was studied numerically and the results qualitatively coincide with those obtained from the linear stability theory. 

These results open new possibilities for experiments on viscous fingering under periodic forcing in the presence of a non-monotonic viscosity profile for immiscible fluids. Another perspective of this work is to perform new numerical simulations of the observed plumes to understand the mechanism that eventually may lead to the interface breaks up.



\section{Acknowledgments}
This research has been supported by the Ministerio de Ciencia e Innovaci\'on (grant no. RTI2018-097063-B-I00) and by Xunta de Galicia (grant no. 2021-PG036). Both programmes are co-funded by ERDF (EU). Computations took place at CESGA (Centro de Supercomputaci\'on de Galicia).


\begin{thebibliography}{99}

\bibitem{Taylor58} G.I. Taylor and P.G. Saffman, Proc. R. Soc. London A \textbf{245}, 312 (1958).
\bibitem{Faber95} T.E. Faber. \emph{Fluid Dynamics for Physicists}. (Cambridge Univ. Press, New York, 1995).
\bibitem{Nittmann85} J. Nittmann, G. Daccord, and H. Stanley, Nature (London) \textbf{314}, 141 (1985); J. Nase, A. Lindner, and C. Creton, Phys. Rev. Lett. \textbf{101}, 074503 (2008).
\bibitem{Wilson90} S.D.R. Wilson, J. Fluid Mech. \textbf{220}, 413--425 (1990).
\bibitem{Mora} S. Mora and M. Manna, Phys. Rev. E \textbf{80}, 016308 (2009); S. Mora and M. Manna, Phys. Rev. E \textbf{81}, 026305 (2010).

\bibitem{Araktingi93} U. Araktingi and F. Orr, SPE Adv. Technol. Series \textbf{1}, 71--80 (1993).
\bibitem{Nagatsu} Y. Nagatsu, K. Matsuda, Y. Kato, and Y. Tada, J. Fluid Mech. \textbf{571}, 475--493 (2007); Y. Nagatsu and A. De Wit, Phys. of Fluids \textbf{23}, 043103 (2011).
\bibitem{DeWit20} A. De Wit, Annu. Rev. Fluid Mech. \textbf{52}, 531--555 (2020).
\bibitem{Escala21} D.M. Escala and A.P. Mu\~nuzuri. Sci. Reports \textbf{11}, 24368 (2021). 
\bibitem{Tafur21} N. Tafur, D.M. Escala, A. Soto, and A.P. Mu\~nuzuri. Chaos \textbf{31}, 023135 (2021). 
\bibitem{Trevelyan11} P.M.J. Trevelyan, C. Almarcha, and A. De Wit, J. Fluid Mech. \textbf{670}, 38--65 (2011).
\bibitem{Riolfo12} L.A. Riolfo, Y. Nagatsu, S. Iwata, R. Maes, P.M.J. Trevelyan, and A. De Wit, Phys. Rev. E \textbf{85}, 015304 (2012).

\bibitem{Escala19} D. M. Escala, A. De Wit, J. Carballido-Landeira, and A.P. Mu\~nuzuri, Langmuir \textbf{35}, 4182--4188 (2019).
\bibitem{Daripa} P. Daripa and G. Paca, Appl. Math. Lett. \textbf{18}, 1293--1303 (2005); P. Daripa, Phys. Fluids \textbf{20}, 112101 (2008); P. Daripa and X. Ding, Transp. Porous Med. \textbf{93}, 675--703 (2012).
\bibitem{Manickam93} O. Manickam, and G. M. Homsy, Phys. Fluids A \textbf{5}, 1356--1367 (1993).
\bibitem{Haudin16} F. Haudin, M. Callewaert, W. De Malsche, and A. De Wit, Phys. Rev. Fluids \textbf{1}, 074001 (2016).


\bibitem{Dias13} E.O. Dias and J.A. Miranda, Phys. Rev. E \textbf{87}, 053015 (2013).
\bibitem{Zheng15} Z. Zheng, H. Kim, and H.A. Stone, Phys. Rev. Lett. \textbf{115}, 174501 (2015).
\bibitem{Pihler18} D. Pihler-Puzivic, G.G. Peng, J.R. Lister, M. Heil, and A. Juel, J. Fluid Mech. \textbf{849}, 163--191 (2018).

\bibitem{Ruith00} M. Ruith and E. Meiburg, J. Fluid Mech. \textbf{420}, 225--257 (2000).

\bibitem{Barrio07} V.L. Barrio, G. Schaub, M. Rohde, S. Rabe, F. Vogel, J.F. Cambra, P.L. Arias, and M.B. Gemez, Intl. J. Hydrogen Energy \textbf{32}, 1421-–1428 (2007).
\bibitem{Torralba08} M. Torralba, J. Ortin, A. Hern\'andez-Machado, and E. Corvera, Phys. Rev. E \textbf{77}, 036207 (2008).
\bibitem{Gosayir13} M. Al-Gosayir, J. Leung, T. Babadagli, and A.M.M. Al- Bahlani, J. Petrol. Sci. Engng \textbf{110}, 74--84 (2013).
\bibitem{Lins16} T.F. Lins and J. Azaiez, Can. J. Chem. Eng. \textbf{94}, 2061--2071 (2016).

\bibitem{Morrow19} L. C. Morrow, T. J. Moroney, and S. W. McCue, J. Fluid Mech. \textbf{877}, 1063--1097 (2019).
\bibitem{Coutinho20} I.M. Coutinho and J.A. Miranda, Phys. Rev. E \textbf{102}, 063102 (2020).

\bibitem{Lins17} S. Li, J.S. Lowengrub, J. Fontana, and P. Palffy-Muhoray, Phys. Rev. Lett. \textbf{102}, 174501 (2016); T.F. Lins and J. Azaiez, J. Fluid Mech. \textbf{819}, 713--729 (2017).
\bibitem{mixing} B. Jha, L. Cueto-Felgueroso, and R. Juanes, Phys. Rev. Lett. \textbf{106}, 194502 (2011);  B. Jha, L. Cueto-Felgueroso, and R. Juanes, Phys. Rev. E \textbf{84},
066312 (2011); C.-Y. Chen, Y.-C. Huang, Y.-S. Huang, and J.A. Miranda Phys. Rev. E \textbf{92}, 043008 (2015).

\bibitem{Chorin68} A.J. Chorin, Mathematics of Computation \textbf{22}, 745--762 (1968).
\bibitem{Moin85} J. Kim and P. Moin, J. of Comp. Phys. \textbf{59}, 308--323 (1985); H. Le and P. Moin, J. of Comp. Phys. \textbf{92}, 369--379 (1991).

\bibitem{interface} A. Elgowainy and N. Ashgriz, Physics of Fluids \textbf{9}, 1635--1649 (1997); R. Scardovelli and S. Zaleski, Annu. Rev. Fluid Mech. \textbf{31}, 567--603 (1999); 
G. Terrones and T. Heberling, Physics of Fluids \textbf{32}, 094105 (2020); H.R. Liu, K. Leong Chong, Q. Wang, C. ShenNg, R. Verzicco, and D. Lohse, J. Fluid Mech. \textbf{913}, A9 (2021). 

\end{thebibliography}
\end{document}